\documentclass[epj,final]{svjour}
\usepackage[usenames,dvipsnames]{color}
\usepackage{graphicx}
\usepackage{multirow}

\DeclareGraphicsExtensions{.jpg}

\newcommand{\secref}[1]{Section~\ref{sec_#1}}
\newcommand{\figref}[1]{Figure~\ref{fig_#1}}
\newcommand{\tblref}[1]{Table~\ref{tbl_#1}}
\newcommand{\eqref}[1]{equation~(\ref{eq_#1})}

\newcommand{\citet}[2]{#2~\cite{#1}}
\newcommand{\etal}[0]{~\textit{et~al.}}

\newcommand{\argmax}[0]{\arg\!\max}

\newcommand{\lpa}[0]{LPA}
\newcommand{\gpa}[0]{GPA}
\newcommand{\gpan}[0]{GPA$_\mathrm{N}$}
\newcommand{\gpac}[0]{GPA$_\mathrm{C}$}
\newcommand{\mmem}[0]{MM$_\mathrm{EM}$}

\newcommand{\jungc}[0]{[\texttt{jung.visualization.}]~\texttt{*(Server|Viewer|Pane|Model|Context)}~(9);~\texttt{cont\-rol.*}~(4);~\texttt{control.*Control}~(5);~\texttt{layout.*}~(7);~\texttt{picking.*State}~(3);~\texttt{pick\-ing.*Support}~(6);~\texttt{renderers.*Renderer}~(13);~\texttt{renderers.*Support}~(3);~etc.}
\newcommand{\jungfstnw}[0]{[\texttt{jung.algorithms.filters.}]~\texttt{*Filter}~(3).}
\newcommand{\jungfstne}[0]{[\texttt{jung.graph.}]~\texttt{*(Graph|Multigraph|Tree)}~(18);~etc.}
\newcommand{\jungfstc}[0]{[\texttt{jung.}]~\texttt{algorithms.generators.*Generator}~(2);~\texttt{algorithms.importance. *}~(4);~\texttt{algorithms.layout.*Layout*}~(3);~\texttt{algorithms.scoring.*Scorer}~(2); \texttt{algorithms.shortestpath.*}~(2);~\texttt{graph.*(Graph|Tree|Forest)}~(4);~etc. (interfaces)}
\newcommand{\jungfstsw}[0]{[\texttt{jung.algorithms.}]~\texttt{layout.*Layout*}~(7);~\texttt{layout3d.*Layout}~(3);~etc.}
\newcommand{\jungfstse}[0]{[\texttt{jung.}]~\texttt{algorithms.cluster.*Clusterer*}~(4);~\texttt{algorithms.generators. random.*Generator}~(5);~\texttt{algorithms.importance.*Betweenness*}~(3);~\texttt{alg\-orithms.metrics.*}~(3);~\texttt{algorithms.scoring.**}~(5);~\texttt{algorithms.short\-estpath.*}~(5);~\texttt{graph.util.*}~(7);~etc.~(implementations)}
\newcommand{\jungsndn}[0]{[\texttt{jung.io.graphml.}]~\texttt{parser.*Parser}~(10);~etc.}
\newcommand{\jungsnds}[0]{[\texttt{jung.io.graphml.}]~\texttt{*Metadata}~(8);~etc.}
\newcommand{\jungrd}[0]{[\texttt{jung.visualization.control.}]~\texttt{*Plugin}~(2).}

\newcommand{\javaxc}[0]{[\texttt{javax.swing.}]~\texttt{plaf.*UI}~(24);~\texttt{plaf.basic.Basic*UI}~(42);~\texttt{plaf.metal.Me\-tal*UI} (22);~\texttt{plaf.multi.Multi*UI}~(30);~\texttt{plaf.synth.Synth*UI}~(40);~etc.}
\newcommand{\javaxn}[0]{[\texttt{javax.}]~\texttt{accessibility.Accessible*}~(10);~\texttt{swing.J*}~(41);~\texttt{swing.**(Bor\-der|Borders|Box|Button|Dialog|Divider|Editor|Factory|Filter|Icon |Kit|LookAndFeel|Listener|Model|Pane|Panel|Popup|Renderer|UIRes\-ource|View)}~(92);~etc.}
\newcommand{\javaxw}[0]{[\texttt{javax.}]~\texttt{accessibility.Accessible*}~(6);~\texttt{swing.*}~(34);~\texttt{swing.event.*Ev\-ent}~(8);~\texttt{swing.event.*Listener}~(13);~\texttt{swing.plaf.*UI}~(6);~etc.}
\newcommand{\javaxs}[0]{[\texttt{javax.swing.}]~\texttt{text.*View}~(15);~\texttt{text.html.*View}~(16);~etc.}

\begin{document}

%%%%%%%%%%%%%%%%%%%%%%%%%%%%%%%%%%%%%%%%%%%%%%%%%%%%%%%%%%%%

\title{Ubiquitousness of link-density and link-pattern communities in real-world networks}
\author{L. \v Subelj\thanks{\email{lovro.subelj@fri.uni-lj.si}} \and M. Bajec}
\institute{University of Ljubljana, Faculty of Computer and Information Science, Ljubljana, Slovenia}
\date{\today}

\abstract{Community structure appears to be an intrinsic property of many complex real-world networks. However, recent work shows that real-world networks reveal even more sophisticated modules than classical cohesive (link-density) communities. In particular, networks can also be naturally partitioned according to similar patterns of connectedness among the nodes, revealing link-pattern communities. We here propose a propagation based algorithm that can extract both link-density and link-pattern communities, without any prior knowledge of the true structure. The algorithm was first validated on different classes of synthetic benchmark networks with community structure, and also on random networks. We have further applied the algorithm to different social, information, technological and biological networks, where it indeed reveals meaningful (composites of) link-density and link-pattern communities. The results thus seem to imply that, similarly as link-density counterparts, link-pattern communities appear ubiquitous in nature and design.}

% 89.75.Fb, 89.75.Hc, 89.75.Kd, 89.20.-a, 89.65.-s

\maketitle

%%%%%%%%%%%%%%%%%%%%%%%%%%%%%%%%%%%%%%%%%%%%%%%%%%%%%%%%%%%%

\section{\label{sec_intro}Introduction}
Complex real-world networks commonly reveal local cohesive modules of nodes denoted (\textit{link-density}) \textit{communities}~\cite{GN02}. These are most frequently observed as densely connected clusters of nodes that are only loosely connected between. Communities possibly play crucial roles in different real-world systems~\cite{GD03,PDFV05}; furthermore, community structure also has a strong impact on dynamic processes taking place on networks~\cite{ADP06,ZGC10}. Thus, communities provide an insight into not only structural organization but also functional behavior of various real-world systems~\cite{PDFV05,RB03,LLDM09,FFGP11}.

Consequently, analysis of community structure is currently considered one of the most prominent areas of network science~\cite{DDDA05,For10,POM09}, while it has also been the focus of recent efforts in a wide variety of other fields. Besides providing many significant theoretical grounds~\cite{FFGP11}, a substantial number of different community detection algorithms has also been proposed in the literature~\cite{WH04,GA05,RB07,SJN06,RAK07,AK08,EL10,Liu10,SB11a} (for reviews see~\cite{For10,POM09,New04a}). However, most of the past research was focused primarily on classical communities characterized by higher density of edges~\cite{RCCLP04}. In contrast to the latter, some recent work demonstrates that real-world networks reveal even more sophisticated communities~\cite{NL07,LXZY07,PSR10,PMB10} that are indistinguishable under classical frameworks.

\footnotetext[1]{Link-pattern communities are known as blockmodels~\cite{WBB76} in social networks literature. These were rigorously analyzed in the past, however, the main focus and employed formulation differs from ours.}

Networks can also be naturally partitioned according to similar patterns of connectedness among the nodes, revealing \textit{link-pattern communities}~\cite{NL07,LXZY07}. (The term was formulated by~\citet{LXZY07}{Long\etal}.) Loosely speaking, link-pattern communities correspond to clusters of nodes that are similarly connected with the rest of the network (i.e., share common neighborhoods). Note that link-density communities are in fact a special case of link-pattern communities (with some fundamental differences discussed later on). Thus, some of the research on the former also apply for the latter~\cite{GA05,RB07,ZLZ11b,LBA11,DKMZ11b}. However, contrary to the flourish of the literature on classical communities in the last decade, a relatively small number of authors have considered more general link-pattern counterparts~\cite{NL07,LXZY07,PSR10,PMB10,ABFX08,BGM10,LKC10,SC10a,HEPF10,CB10,DKMZ11c} (in the same sense as in this paper\footnotemark[1]). Although this could be attributed to a number of factors like increased complexity or lack of adequate generative models and algorithms, more importantly, existence of meaningful link-pattern communities has not been properly verified under the same framework in various different types of real-world networks that are commonly analyzed in the literature (still, some networks have been considered in the past). In this paper we try to address this issue. (Note that similar stance was also made by~\citet{NL07}{Newman~and~Leicht}.)

We extend balanced propagation~\cite{SB11c} with defensive preservation of communities~\cite{SB11a} into a general approach that can extract arbitrary network modules ranging from link-density to link-pattern communities. To the best of our knowledge, this is the only such algorithm that does not require some prior knowledge of the true structure (e.g., the number of communities), or does not optimize some heuristic selected beforehand. We have validated the proposed algorithm on two classes of synthetic benchmark networks with community structure, and also on random networks. The algorithm was further applied to different social, information, technological and biological networks, where it reveals meaningful composites of link-density and link-pattern communities that are well supported by the network topology. The results thus seem to imply that, similarly as link-density counterparts, link-pattern communities appear ubiquitous in nature and technology.

The rest of the paper is structured as follows. In~\secref{ldpc} we discuss the relation between link-density and link-pattern communities in greater detail, and propose a propagation based algorithm for their detection. Results on synthetic and real-world networks are presented and formally discussed in~\secref{res}, while in~\secref{conc} we summarize our main observations and discuss some prominent directions for future research.

%%%%%%%%%%%%%%%%%%%%%%%%%%%%%%%

\section{\label{sec_ldpc}Link-density and link-pattern communities}
\begin{figure}[t]
\centering
\includegraphics[width=1.00\columnwidth]{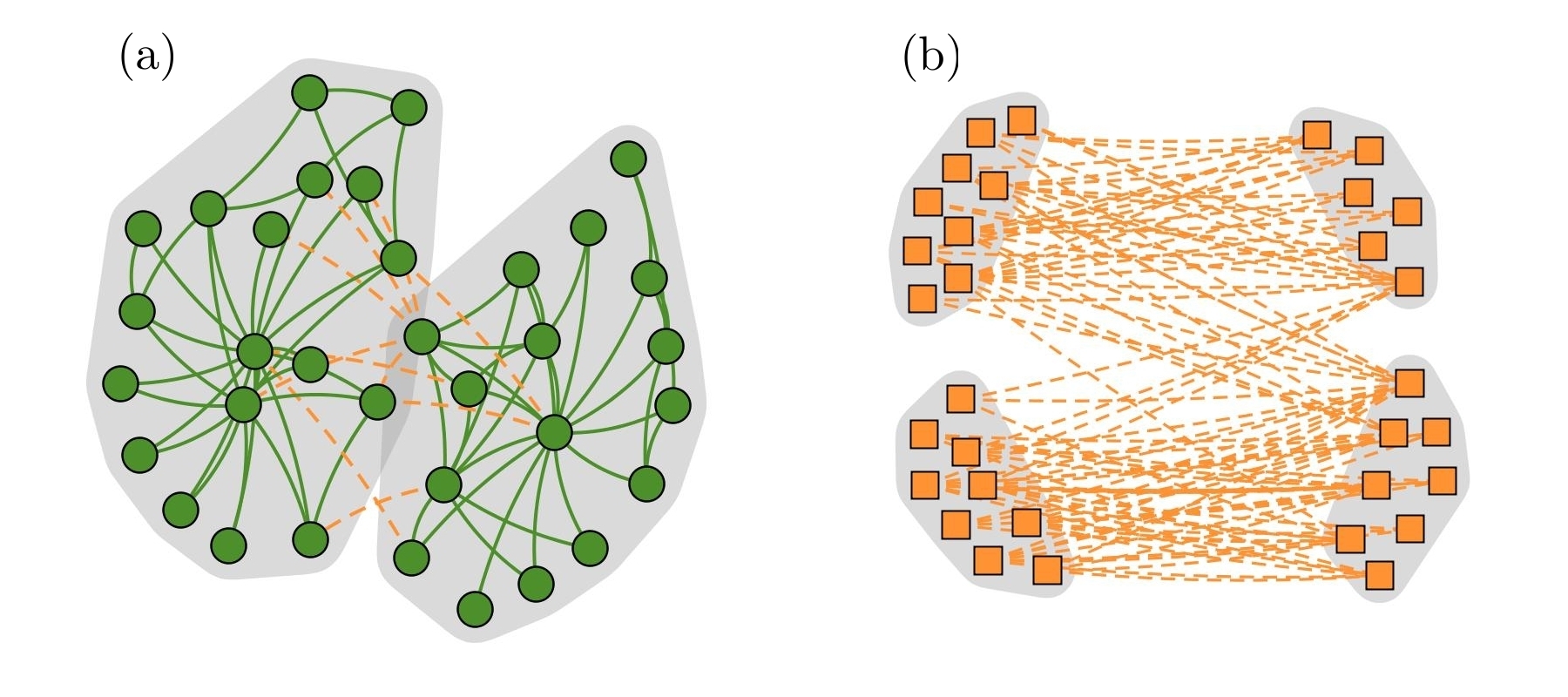} % 40 pt
\caption{\label{fig_ldpc}(Color~online)~Link-density and link-pattern communities (i.e., shaded regions) in (a)~\textit{karate} and (b)~\textit{women} social networks, respectively (\tblref{resrw}). The former represents social interactions among the members of karate club observed by~\citet{Zac77}{Zachary}, while the latter shows social events (right-hand side) visited by women (left-hand side) in Natchez, Mississippi~\cite{DGG41}. Link-density communities correspond to cohesive modules of nodes, whereas link-pattern communities represent common patterns of connectedness among the nodes.}
\end{figure}

Although classical link-density communities can be considered under the same framework as link-pattern communities, there exist some significant differences between the two (\figref{ldpc}). Most obviously, link-pattern communities do not correspond to cohesive modules of nodes, whereas, such communities commonly do not even feature connectedness. Connectedness is considered a fundamental structural property of link-density communities, and thus a common ingredient of different objective functions and community detection algorithms~\cite{For10}.

While link-density communities are often related to the notions of assortative mixing by degree and homophily~\cite{GN02,NG03} (at least in social networks), link-pattern communities might in fact represent the origin of more commonly observed disassortative degree mixing~\cite{New02,NP03}. As the latter has been analyzed to much lesser extent than the former~\cite{NL07}, direct dependence has not yet been verified in real-world networks. Nevertheless, disassortative mixing refers merely to the phenomena that nodes mainly connect to dissimilar nodes, and thus outside of their respective (link-pattern) community. However, how such communities relate between each other, and with the rest of the network, remains unexplained. (Although network assortativity was most commonly analyzed in the context of node degree~\cite{New02,NG03}, we refer to the notion in general.)

Note also that, as nodes of some link-pattern community are commonly not directly connected, they exhibit somewhat higher mutual independence than nodes within some link-density community. On the contrary, nodes from neighboring link-pattern communities are somewhat more dependent than in the case of classical communities.

Due to all above, we strictly distinguish between link-density and link-pattern communities within the proposed algorithm. However, it ought to be mentioned that this is rather different from other authors, who have typically considered all communities under link-pattern regime~\cite{NL07,LXZY07,PSR10,PMB10,ABFX08,BGM10,LKC10}. Nevertheless, the latter could be attributed to the fact that other approaches are mainly based on previous work in social sciences, statistics or artificial intelligence, where such setting might be more adequate.

In~\secref{ldpc_bpa} we first introduce a balanced propagation based algorithm for classical community detection; while the algorithm is extended for general community detection in~\secref{ldpc_gpa}.

%%%%%%%%%%%%%%%%%%%%%%

\subsection{\label{sec_ldpc_bpa}Classical community detection}
Let the network be represented by an undirected and unweighted multi-graph $G(N,E)$, with $N$ being the set of nodes of the graph and $E$ being the set of edges. Furthermore, let $c_n$ be the community (label) of node $n$, $n\in N$, and $\mathcal{N}(n)$ the set of its neighbors.

Algorithms presented below are in fact based on a label propagation proposed by~\citet{RAK07}{Raghavan\etal}. The label propagation algorithm (\lpa)~\cite{RAK07} extracts (link-density) communities by exploiting the following simple procedure. At first, each node is labeled with a unique label, $c_n=l_n$. Then, at each iteration, each node adopts the label shared by most of its neighbors. Hence,
\begin{eqnarray}
\label{eq_lpa}
c_n & = & \argmax_l|\mathcal{N}^l(n)|,
\end{eqnarray}
where $\mathcal{N}^l(n)$ is the set of neighbors of $n$ that share label~$l$ (ties are broken uniformly at random). To prevent oscillations of labels, node $n$ retains its current label when it is among most frequent in $\mathcal{N}(n)$~\cite{RAK07}. Due to existence of many intra-community edges, relative to the number of inter-community edges, nodes in a (link-density) community form a consensus on some particular label after a few iterations. Thus, when an equilibrium is reached, disconnected groups of nodes sharing the same label are classified into the same community. 

Due to extremely fast structural inference of label propagation, the algorithm exhibits near linear time complexity~\cite{RAK07,SB11a} (in the number of edges) and can easily scale to networks with millions (or even billions) of nodes and edges~\cite{SB11a,SB10a}. Also, due to its algorithmic simplicity, it is currently among more commonly adopted algorithms in the literature. Still, label propagation can be further improved in various ways~\cite{BC09a,LHLC09,Gre10,BRSV11,SB11c,SB11a}.

In the following we present two advances of the basic approach that improve on its robustness and community detection strength. Both result in a simple incorporation of node preferences~$p_n$~\cite{LHLC09} into the updating rule of label propagation as
\begin{eqnarray}
\label{eq_ppa}
c_n & = & \argmax_l\sum_{m\in\mathcal{N}^l(n)}p_m.
\end{eqnarray}
 (See \eqref{dbpa}.) Node preferences adjust the propagation strength of each respective node, and can thus direct the propagation process towards more desirable partitions~\cite{LHLC09,SB11a}. Note that preferences $p_n$ can be set to an arbitrary node statistic (e.g., degree~\cite{LHLC09}). 

To address issues with oscillations of labels in some networks (e.g., bipartite networks), nodes' labels are updated in a random order~\cite{RAK07} (independently among iterations). Although this solves the aforementioned problem, the introduction of randomness severely hampers the robustness of the algorithm, and consequently also the stability of the identified community structure. Different authors have noted that label propagation reveals a large number of different community structures even in smaller networks, while these structures are also relatively different among themselves~\cite{TK08,SB11a}.

We have previously shown that updating nodes in some particular order results in higher propagation strength for nodes that are updated at the beginning, and lower propagation strength for nodes that are updated towards the end~\cite{SB11c}. The order of node updates thus governs the algorithm in a similar manner than corresponding node propagation preferences. Based on the latter, we have proposed a balanced propagation algorithm~\cite{SB11c} that utilizes node preferences to counteract (i.e., balance) the randomness introduced by random update orders. In particular, we introduce the notion of node balancers that are set to the reverse order in which the nodes are assessed. Thus, lower and higher propagation strength is assigned to nodes considered first and last, respectively.

Let nodes $N$ be ordered in some random way, and let $i_n$ denote the normalized position of node $n$ in this order, $i_n\in(0,1]$. Hence,
\begin{eqnarray}
\label{eq_ind}
i_n & = & \frac{\mbox{index of node }n}{|N|}.
\end{eqnarray}
Node balancers $b_n$ can then be modeled with a simple linear function as $b_n=i_n$. However, using a logistic curve allows for some further control over the algorithm. Thus,
\begin{eqnarray}
\label{eq_bp}
b_n & = & \frac{1}{1+\exp(-\beta(i_n-\alpha))},
\end{eqnarray}
where $\alpha$ and $\beta$ are parameters of the algorithm. Intuitively, we fix $\alpha$ to $0.5$, while stability parameter $\beta$ is set to $0.25$ according to some preliminary experiments~\cite{SB11c} (see below). Note that balancers $b_n$ are re-estimated before each iteration, and are incorporated into the algorithm as node propagation preferences (see~\eqref{dbpa}). 

Setting the stability parameter $\beta$ to $0$ yields the basic label propagation approach, while increasing $\beta$ significantly improves the robustness of the algorithm. However, computational complexity thus also increases. Hence, balanced propagation improves the stability of the identified community structure for the sake of higher complexity, while the trade-off is in fact governed by the parameter $\beta$. Note that community detection strength of the refined algorithm is also improved in most cases. For a more detailed discussion see~\cite{SB11c}.

To even further improve the performance of the algorithm we also adopt defensive preservation of communities~\cite{SB11a}. The strategy increases the propagation strength from the core of each currently forming community, which results in an immense ability of detecting communities, even when they are only weakly depicted in the network's topology. Laying the pressure from the borders also prevents a single community from occupying a large portion of the network, which else occurs in, e.g., information networks~\cite{LHLC09}. Thus, the strategy defensively preserves communities and forces the algorithm to more gradually reveal the final structure. For further discussion see~\cite{SB11a,SB10a}.

In the algorithm, community cores are estimated by means of the diffusion over the network. The latter is modeled by employing a random walker within each community. Let $d_n$ be the probability that a random walker utilized on community $c_n$ visits node $n$. Then,
\begin{eqnarray}
\label{eq_dp}
d_n & = & \sum_{m\in\mathcal{N}^{c_n}(n)}\frac{d_m}{k_m^{c_n}},
\end{eqnarray}
where $k_m^{c_n}$ is the intra-community degree of node $m$. Besides deriving the estimates of cores and borders, the main objection here is to mimic label propagation within each community, to estimate the current state of the propagation process, and then to adequately alter its dynamics (see~\eqref{dbpa}). Note that values $d_n$ are re-estimated according to~\eqref{dp} when the corresponding node updates its label (initially, all $d_n$ are set to $1/|N|$).

Similarly as above, diffusion values $d_n$ are incorporated into the algorithm as node propagation preferences. Thus, the updating rule for balanced propagation algorithm with defensive preservation of communities is
\begin{eqnarray}
\label{eq_dbpa}
c_n & = & \argmax_l\sum_{m\in\mathcal{N}^l(n)}b_md_m.
\end{eqnarray}

The above is taken as a basis for a general community detection algorithm presented in the following section. Note that the formulation can be extended to weighted networks in a straightforward fashion.

%%%%%%%%%%%%%%%%%%%%%%

\subsection{\label{sec_ldpc_gpa}General community detection}
Label propagation algorithm (and its advances) cannot be directly adopted for detection of link-pattern communities, as the bare nature of label propagation demands cohesive (connected) clusters of nodes (\secref{ldpc}). However, link-pattern communities can still be seen as cohesive modules when one considers second order neighborhoods (i.e., nodes at distance $2$). Thus, instead of propagating labels between the neighboring nodes, the labels are rather propagated through node's neighbors (i.e., between nodes at distance $2$). For instance, when a group of nodes exhibits similar pattern of connectedness with other nodes, propagating labels through these latter nodes would indeed reveal the respective link-pattern community (similarly as for classical label propagation).

Considering the above, balanced propagation based algorithm presented in~\secref{ldpc_bpa} can be extended for link-pattern communities in a rather \textit{ad hoc} fashion. Let $\delta_l$ be a community dependent parameter, $\delta_l\in[0,1]$, such that $\delta_l\approx 1$ and $\delta_l\approx 0$ for link-density and link-pattern communities, respectively. Thus, when $\delta_l$ varies from $1$ to $0$, communities range from classical link-density communities to proper link-pattern communities. Balanced propagation in~\eqref{dbpa} can then be simply advanced into a general community detection algorithm as
\begin{eqnarray}
\label{eq_gpa}
c_n & = & \argmax_l\left(\delta_l\sum_{m\in\mathcal{N}^l(n)}b_md_m\right. + \\
\nonumber
 & & + \left.(1-\delta_l)\sum_{m\in\mathcal{N}^l(s)|s\in\mathcal{N}(n)}b_m\frac{d'_m}{k_s}\right),
\end{eqnarray}
where similarly as in~\eqref{dp}, diffusion values $d'_n$ are estimated using random walks. Hence,
\begin{eqnarray}
\label{eq_dp2}
d'_n & = & \sum_{m\in\mathcal{N}^{c_n}(s)|s\in\mathcal{N}(n)}\frac{d'_m}{\sum_{s\in\mathcal{N}(m)}k_{s}^{c_n}}.
\end{eqnarray}
(Denominators in~equations~(\ref{eq_gpa}),~(\ref{eq_dp2}) provide that the sums are proportional to the degree of the node $k_n$.) Else, the proposed algorithm is identical as before, and is denoted general propagation algorithm (\gpa). Note that setting all $\delta_l$ to $1$ yields the classical community detection algorithm in~\eqref{dbpa}.

Due to simplicity, in \gpa~all $\delta_l$ are fixed to $0.5$. Still, the algorithm can detect either link-density or link-pattern communities, or different mixtures of both, when they are clearly depicted in the network's topology (\secref{res}). However, the algorithm can also detect communities that are of neither link-density nor link-pattern type.

As our main intention is to unfold meaningful composites of mainly link-density and link-pattern communities, we also propose a variant of the algorithm denoted \gpac. The latter algorithm re-estimates the values $\delta_l$ on each iteration, in order to reveal clearer community structure. In particular, we measure the quality of each community using the conductance~$\Phi$~\cite{Bol98}, to determine whether the community better conforms with link-density or link-pattern regime. (The conductance measures the goodness of a link-density community, or equivalently, the quality of the corresponding network cut.)  As good link-density communities exhibit low values of conductance, and good link-pattern communities exhibit high values, after each iteration of the algorithm (though omitted on first) we set $\delta_l$ according to
\begin{eqnarray}
\label{eq_cdel}
\delta_l & = & 1-\Phi(l) = \frac{1}{k^l}\sum_{n\in N^l}k^{l}_n,
\end{eqnarray}
where $k^l$ is the strength of community $l$, $k^l=\sum_{n\in N^l}k_n$ (initially all $\delta_l$ are set to $0.5$). As the strategy adjusts values of $\delta_l$ with respect to each individual community, the algorithm more accurately reveals different composites of link-density and link-pattern communities (\secref{res}).

For networks with clear assortative or disassortative mixing, values $\delta_l$ can in fact be more accurately estimated on the level of the entire network (\secref{res}). Hence, 
\begin{eqnarray}
\label{eq_ndel}
\delta_l & = & \sum_l\frac{|N^l|}{|N|}(1-\Phi(l)), %= \\
%\nonumber
% & & = \frac{1}{|N|}\sum_l\frac{|N^l|}{k^l}\left(\sum_{n\in N^l}k^{l}_n\right),
\end{eqnarray}
while the resulting algorithm is denoted \gpan.

All proposed algorithms have complexity near $O(k|E|)$, where $k$ is the average degree in the network. 

%%%%%%%%%%%%%%%%%%%%%%%%%%%%%%%

\section{\label{sec_res}Results and discussion}
In the following sections we analyze the proposed algorithms on different synthetic and real-world networks (\secref{ressyn} and \secref{resrw}, respectively).

General propagation algorithms (i.e., \gpa, \gpac~and \gpan) are compared against two other approaches. As a representative of classical community detection algorithms, we employ basic label propagation (i.e., \lpa). Next, we also adopt the mixture model with expectation-maximization~\cite{DLR77} proposed by~\citet{NL07}{Newman~and~Leicht} (denoted \mmem). Their algorithm can detect arbitrary network modules and is currently among state-of-the-art approaches for detection of link-pattern communities~\cite{NL07,PSR10}. Still, it demands the number of communities to be known beforehand. Note that the exact number of communities (currently) cannot be adequately estimated in large real-world networks~\cite{KN11a}. Due to simplicity, we limit the number of iterations to $100$ for all the algorithms.

The results are assessed in terms of normalized mutual information (NMI)~\cite{DDDA05}, which has become a \textit{de facto} standard in community detection literature. Let $\mathcal{C}$ be a partition revealed by the algorithm and let $\mathcal{P}$ be the true partition of the network (corresponding random variables are $C$ and $P$, respectively). NMI of $\mathcal{C}$ and $\mathcal{P}$ is then
\begin{eqnarray}
\mbox{NMI} & = & \frac{2I(C,P)}{H(C)+H(P)},
\label{eq_nmi}
\end{eqnarray}
where $I(C,P)$ is the mutual information of the partitions, i.e., $I(C,P)=H(C)-H(C|P)$, and $H(C)$, $H(P)$ and $H(C|P)$ are standard and conditional entropies. NMI of identical partitions equals $1$, and is $0$ for independent ones.

%%%%%%%%%%%%%%%%%%%%%%

\subsection{\label{sec_ressyn}Synthetic networks}
The algorithms were first applied to synthetic benchmark networks with two communities of $32$ nodes. Average degree is fixed to $6$, while the community structure is controlled by a mixing parameter $\mu$, $\mu\in[0,1]$. When $\mu$ equals $0$, all edges are (randomly) placed between the nodes of the same community, and when $\mu$ equals $1$, all edges are (randomly) placed between the nodes of different communities. Thus, when  $\mu$ varies from $0$ to $1$, community structure ranges between link-density and link-pattern regime (i.e., assortative and disassortative mixing). Note that network structure is completely random for $\mu=0.5$.

\begin{figure}[t]
\centering
\includegraphics[width=0.80\columnwidth]{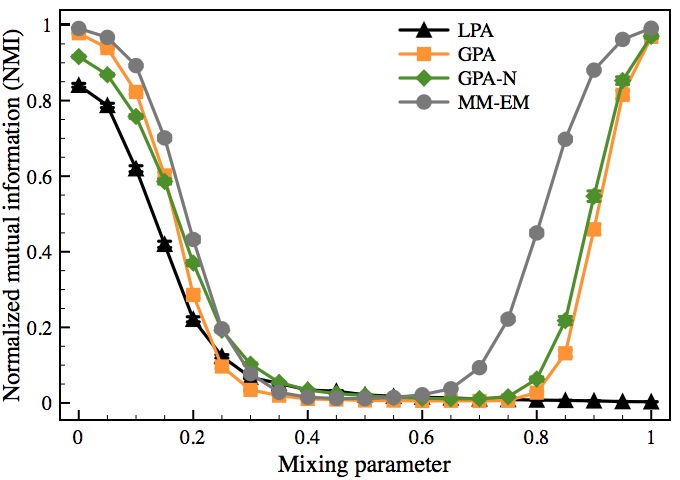} % |N|=2*32, <k>=6
\caption{\label{fig_ressyn64}(Color~online)~Mean NMI over $1000$ realizations of synthetic networks with two communities. Error bars showing standard error of the mean are smaller than the symbol sizes.}
\end{figure}

The results appear in~\figref{ressyn64}. As anticipated, classical community detection algorithm \lpa~is unable to distinguish between a network with disassortative mixing and a completely random network (i.e., $\mu\approx 1$ and $\mu\approx 0.5$, respectively). Moreover, \lpa~also has the worst performance for all community regimes. On the other hand, mixture model \mmem~performs significantly better than other algorithms, especially in the case of link-pattern communities (i.e., $\mu>0.5$). However, we argue that this is largely due to the fact that the algorithm is given the true number of communities in advance.

Observe that general propagation algorithms \gpa~and \gpan~can indeed detect both link-density and link-pattern communities. However, the algorithm with a network-wise re-estimation of $\delta_l$ performs slightly better, except when the structure results in clear link-density communities (i.e.,  $\mu\leq 0.1$). Still, the analysis on real-world networks in~\secref{resrw} confirms that \gpan~more accurately reveals different types of communities (including link-density).

We further apply the algorithms to a class of benchmark networks also adopted in~\cite{PSR10}. The latter is in fact a generalization of the benchmark proposed by~\citet{GN02}{Girvan~and~Newman} for classical community detection. More precisely, networks comprise four communities of $32$ nodes, thus, two communities correspond to classical link-density modules, while the other two form a bipartite structure of link-pattern communities. The networks are thus neither assortative nor disassortative (but locally assortative or disassortative). Average degree is fixed to $16$, while the community structure is again controlled by a mixing parameter $\mu$, $\mu\in[0,1]$. Lower values correspond to clearer community structure---when $\mu=0.5$, one half of the edges is set according to the designed structure, while the other are placed at random (on average).

\begin{figure}[b]
\centering
\includegraphics[width=0.80\columnwidth]{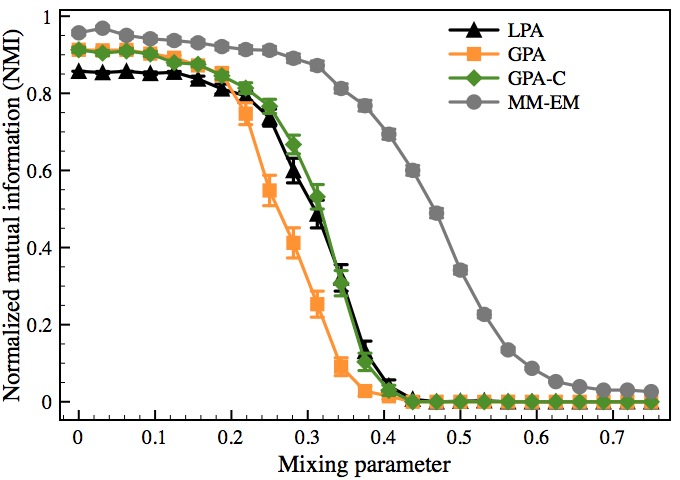} % |N|=4*32, <k>=16
\caption{\label{fig_ressyn128}(Color~online)~Mean NMI over $100$ realizations of synthetic networks with four communities. Error bars show standard error of the mean.}
\end{figure}

The results in~\figref{ressyn128} also report the performance of \lpa, although a classical community detection algorithm is obviously not suited for these networks. However, one can thus observe that, when community structure is rather clearly defined (i.e., $\mu<0.25$), only a small improvement can be achieved with a general community detection algorithm (on these networks). Therefore, to more accurately estimate the performance of \gpa~and \gpac, we increase the value of parameter $\beta$ to $4$ (\secref{ldpc_bpa}). This further stabilizes the community structure identified by the algorithms, however, the computational time thus increases.

Mixture model \mmem~performs significantly better than other algorithms, which could be attributed to a known number of communities as above. Otherwise, general propagation algorithms \gpa~and \gpac~both detect link-density and link-pattern communities within these networks, however, only until communities are clearly depicted in the networks' topologies (i.e., $\mu<0.25$). When $\mu$ further increases, the algorithm with a cluster-wise re-estimation of $\delta_l$ still manages to reveal (link-density) communities to some extent, whereas, \gpa~already fails.

Considering also the results reported in~\cite{PSR10}, image graph approach of~\citet{PSR10}{Pinkert\etal} performs even slightly better than \mmem, while the model selection of~\citet{RB07}{Rosvall~and~Bergstrom} is a bit worse than \gpac. Thus, we conclude that general propagation algorithms can indeed reveal link-density and link-pattern communities under the same framework, still, the accuracy on these networks is worse with respect to some other state-of-the-art approaches. However, all these approaches demand the number of communities to be given apriori, thus, the algorithms are actually not fully comparable. Moreover, analysis on real-world networks in~\secref{resrw} reveals that, when the number of communities increases, the above advantage is in fact rendered useless.

\begin{table*}[t]
\centering
\caption{\label{tbl_resrwsoc}~Mean NMI over $10000$ and $1000$ runs for \textit{karate}, \textit{women} and \textit{football}, \textit{corporate} networks, respectively.}
\begin{tabular}{ccccccc}
\hline\noalign{\smallskip}
Network & Communities & \lpa &  \gpa & \gpan & \gpac & \mmem \\
\noalign{\smallskip}\hline\noalign{\smallskip}
\textit{karate} & $2$ & $\mathit{0.6501}$ & $0.6992$ & $0.7625$ & $0.7547$ & $\mathbf{0.7806}$ \\
\textit{football} & $12$ & $\mathit{0.8908}$ & $0.8464$ & $\mathbf{0.8570}$ & $0.8493$ & $0.8069$ \\
\noalign{\smallskip}
\textit{women} & $4$ & - & $0.7663$ & $0.7680$ & $0.7675$ & $\mathbf{0.8337}$ \\ % 0.2996
\textit{corporate} & $8$ ($9$) & - & $0.6680$ & $\mathbf{0.6735}$ & $0.6651$ & $0.5995$ \\ % 0.7212
\noalign{\smallskip}\hline
\end{tabular}
\end{table*}

Note that the above benchmark networks represent a relatively poor description of real-world network structure (see~\secref{resrw}). However, construction of networks with both assortative and disassortative mixing is not straightforward, as one inevitably has to define how link-pattern communities connect with the rest of the network. Still, generalization of hierarchical network model~\cite{RB03,CMN08} appears as the most prominent formulation of different community regimes. Here, probabilities assigned to nodes of a predefined hierarchy of communities dictate the connections between the nodes in the network. High probabilities at the bottom level of the hierarchy yield classical cohesive modules, whereas link-pattern communities are characterized by higher probabilities at one level above.

To further validate the proposition, we have also applied the propagation algorithms to a random graph \`{a}~la Erd\"{o}s-R\'{e}nyi~\cite{ER59} that (presumably) has no community structure. The number of nodes is fixed to $256$, while we vary the average degree $k$ between $2$ and $64$. When $k$ exceeds a certain threshold, all algorithms reveal only trivial communities (i.e., connected components of the network). The transition occurs at $k\approx 8$, $k\approx 10$ and $k\approx 12$ for \lpa, \gpa~and~\gpac, and \gpan, respectively. Hence, community structures revealed by general propagation algorithms are beyond simple random configurations, while the algorithms are also not attributed to resolution limit~\cite{FB07} issues (i.e., existence of an intrinsic scale, below which communities are no longer recognized).

%%%%%%%%%%%%%%%%%%%%%%

\subsection{\label{sec_resrw}Real-world networks}
The proposed algorithms were further applied to ten real-world networks with community structure (\tblref{resrw}). All these networks are commonly analyzed in the community detection literature and include different social, technological, information and biological networks (detailed description is omitted). Due to simplicity, all networks are treated as unweighted and undirected. Furthermore, \textit{corporate}, \textit{jung} and \textit{javax} networks are reduced to largest connected components and treated as simple graphs.

\begin{table}[b]
\centering
\caption{\label{tbl_resrw}~Real-world networks with community structure.}
\begin{tabular}{cccc}
\hline\noalign{\smallskip}
Network & Description & Nodes & Edges \\
\noalign{\smallskip}\hline\noalign{\smallskip}
\textit{karate} & Zachary's karate club.~\cite{Zac77} & $34$ & $78$ \\
\textit{football} & Amer. football league.~\cite{GN02} & $115$ & $616$ \\
\noalign{\smallskip}
\textit{women} & Davis's south. women.~\cite{DGG41} & $18$, $14$ & $89$ \\
\textit{corporate} & Scottish corporates.~\cite{SH80} & $131$, $86$ & $348$ \\
\noalign{\smallskip}
\textit{jung} & JUNG graph library.~\cite{SB11d} & $305$ & $710$ \\
\textit{javax} & Java library (\texttt{javax}).~\cite{SB11d} & $705$ & $3313$ \\
\noalign{\smallskip}
\textit{amazon} & Amazon web graph.~\cite{Aut00f} & $2879$ & $5037$ \\
\noalign{\smallskip}
\textit{protein} & S. cerevisiae proteins.~\cite{PDFV05} & $2445$ & $6265$ \\ % Saccharomyces
\noalign{\smallskip}
\textit{gnutella} & Gnutella peer-to-peer.~\cite{LKF07} & $62586$ & $147892$ \\
\noalign{\smallskip}
\textit{condmatt} & Cond. Matt. archive.~\cite{New01b} & $36458$ & $171736$ \\ 
\noalign{\smallskip}\hline
\end{tabular}
\end{table}

We first consider four well known social networks, name\-ly, \textit{karate}, \textit{football}, \textit{women} and \textit{corporate} networks. The former two represent classical benchmarks for link-density community detection, as they reveal clear assortative mixing (\figref{ldpc},~(a)). On the other hand, the latter two are in fact bipartite networks, thus, the respective network communities can be considered of pure link-pattern type (\figref{ldpc},~(b)). However, the networks are not properly disassortative, due to different types of nodes.

All these networks have known sociological partitions into communities that result from earlier studies, while partition of \textit{corporate} network is limited to only $86$~corporate nodes. Comparison between community structures extracted by different algorithms and known network structures can be seen in~\tblref{resrwsoc}. The number of communities in \mmem~algorithm is set to the true value for all networks except \textit{corporate}, where we set it to nine (\tblref{resrwsoc}).

Although the mixture model \mmem~performs better than general propagation algorithms on synthetic benchmark networks (\secref{ressyn}), the latter appears to be dependent on the number of communities. When the number of communities, and thus the size of the network, is relatively small (i.e., \textit{karate} and \textit{women} networks), the  \mmem~most accurately reveals the true network structure. However, when the number of communities increases (i.e., \textit{football} and \textit{corporate} networks), all propagation algorithms significantly outperform \mmem. The latter can be related to previously discussed weakness of \mmem.

Note that somewhat lower performance of propagation algorithms on \textit{karate} and \textit{women} networks is actually due to the fact that the algorithms reveal three communities in these networks, which does not coincide with the sociological partitioning of the nodes. In particular, the algorithms extract a small module from the larger community in \textit{karate} network (\figref{ldpc},~(a)), and merge the two communities representing events in \textit{women} network (\figref{ldpc},~(b)). However, similarly as in the case of sociological communities, both these structures are well supported by the networks' topologies and thus commonly reported by community detection algorithms in the literature. Considering the partition of \textit{women} network with three communities, \gpa,~\gpan~and \gpac~reveal structures with NMI equal to $0.8769$, $0.8809$ and $0.8799$, respectively, while \mmem~obtains only $0.8027$ (on average).

\begin{figure*}[t]
\centering
\includegraphics[width=1.60\columnwidth]{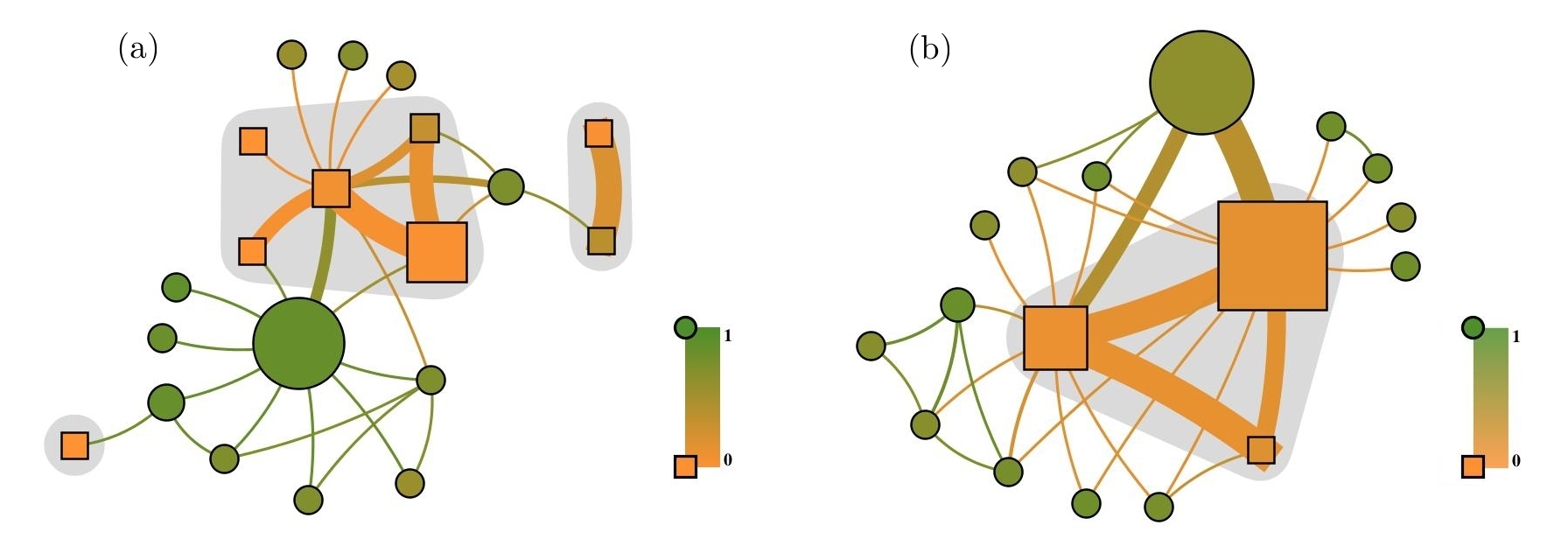} % 28 pt
\caption{\label{fig_resrwtech}(Color~online)~Community structures of (a) \textit{jung} and (b) \textit{javax} technological networks revealed with \gpac. Node sizes are proportional to the community sizes, while the symbols (colors) correspond to the values of $\delta_l$ (\eqref{cdel}).}
\end{figure*}

General propagation algorithms with re-esti\-mation of $\delta_l$, i.e., \gpac~and \gpan, mostly outperform the basic \gpa. As the algorithms adopt to either assortative or disassortative mixing regime in each network, they manage to extract the true communities more accurately. Observe also that network-wise re-estimation is somewhat more adequate for these networks than a cluster-wise version, due to a clear mixing regime. However, for networks with both types of mixing, \gpac~should obviously be employed.

We conclude that general propagation can reveal link-density and link-pattern communities in real-world networks. Thus, exactly the same algorithm is suitable for classical community detection in unipartite networks and link-pattern community detection in multi-partite networks. With respect to high values of NMI in~\tblref{resrwsoc} (except for \textit{corporate} network), the proposed algorithms can also be considered as relatively accurate.

As the above social networks are particularly homogeneous, they reveal either assortative or disassortative mixing. However, social networks could indeed comprise both regimes, still, such networks would have to be heterogeneous by nature (i.e., convey different types of relations between individuals). In fact, heterogeneity seems to be a necessary condition for a network to reveal different composites of link-density and link-pattern communities. In the following we analyze four of the remaining networks in~\tblref{resrw} that are all heterogeneous by nature.

Our main intention in the following is to reveal meaningful composites of not only link-density but also link-pattern communities, and thus imply that such structures could appear ubiquitous in various complex networks. Therefore, we apply \gpac~to each network $10$ times, and report the structure with the highest fraction of nodes within link-pattern communities. It should be noted that community structures of these networks should not be considered identified, as networks possibly reveal a large number of different structures that are all significant and well supported by their topologies~\cite{CMN08} (e.g., communities exist on different scales). Note that multiple structures could also imply that no clear one exists (e.g., overlapping communities~\cite{PDFV05}). However, general propagation algorithms find no communities in random networks (\secref{ressyn}), thus, all revealed structures are at least beyond random.

First, we analyze two technological networks, namely, \textit{jung} and \textit{javax} networks (\tblref{resrw}). These are class dependency networks, where nodes correspond to software classes and edges represent different types of dependencies among them (e.g., inheritance, parameters, variables etc.). The networks are thus obviously heterogeneous and should comprise different types of communities~\cite{SB11d}. 

Revealed community structures are shown in~\figref{resrwtech}. Observe that networks convey both clear link-density and link-pattern communities, whereas, the latter are further combined in rather complex configurations (i.e., shaded regions in~\figref{resrwtech}). In particular, besides simple bipartite structures and isolated link-pattern communities, networks also reveal connected clusters of multiple link-pattern communities. Note that, although link-pattern communities are mainly connected between themselves, they can also be strongly connected with else cohesive modules of nodes. Moreover, both link-density and link-patter communities can reside in either network interior or periphery.

We next analyze the main communities in greater detail (\tblref{resrwtech}). The core, i.e., major link-density community, of \textit{jung} network (\figref{resrwtech},~(a)) consists of only visualization classes, while these are else almost inexistent in other communities. As one could anticipate, the community is highly cohesive and independent from the rest of the network. Two link-pattern communities on the right-hand side contain utility classes for GraphML format; while the upper community mainly contains different par\-sers, the lower mostly consists of meta-data classes, used by the former. Thus, the number of inter-community edges is obviously high. Central configuration of five link-pattern communities also contains well defined modules with particularly clear functional roles. More precisely, communities contain basic graph classes, interfaces for various algorithms, their implementations, different layout classes and filters, respectively. The strength of connections among the communities further supports this functional differentiation (e.g., implementations of different algorithms are strongly dependent on various interfaces and graph classes).

\begin{table*}[t]
\centering
\caption{\label{tbl_resrwtech}~Analysis of community structures revealed in technological networks (\figref{resrwtech}). 'core' denotes the largest link-density community, while '$k$-configuration'-s represent shaded regions in~\figref{resrwtech} ($k$ is the number of link-pattern communities).}
\begin{tabular}{ccccp{10.50cm}}
\hline\noalign{\smallskip}
Network & Community $l$ & $|N^l|$ & $\delta_l$ & \multicolumn{1}{c}{Description} \\
\noalign{\smallskip}\hline\noalign{\smallskip}
\multirow{16}{*}{\textit{jung}} & core & $65$ & $0.86$ & \jungc \\
\noalign{\smallskip}
%\noalign{\smallskip}\cline{2-5}\noalign{\smallskip}
 & 5-conf. (upper left) & $3$ & $0.00$ & \jungfstnw \\
 & 5-conf. (upper right) & $21$ & $0.33$ & \jungfstne \\
 & 5-conf. (central) & $28$ & $0.07$ & \jungfstc \\
 & 5-conf. (lower left) & $13$ & $0.00$ & \jungfstsw \\
 & 5-conf. (lower right) & $44$ & $0.03$ & \jungfstse \\
\noalign{\smallskip}
%\noalign{\smallskip}\cline{2-5}\noalign{\smallskip}
 & 2-conf. (upper) & $13$ & $0.03$ & \jungsndn \\
 & 2-conf. (lower) & $13$ & $0.38$ & \jungsnds \\
\noalign{\smallskip}
%\noalign{\smallskip}\cline{2-5}\noalign{\smallskip}
 & 1-conf. (central) & $2$ & $0.00$ & \jungrd \\
\noalign{\smallskip}\hline\noalign{\smallskip}
\multirow{8}{*}{\textit{javax}} & core & $179$ & $0.64$ & \javaxc \\
\noalign{\smallskip}
%\noalign{\smallskip}\cline{2-5}\noalign{\smallskip}
 & 3-conf. (upper) & $193$ & $0.15$ & \javaxn \\
 & 3-conf. (left) & $113$ & $0.11$ & \javaxw \\
 & 3-conf. (lower) & $44$ & $0.19$ & \javaxs \\
\noalign{\smallskip}\hline
\end{tabular}
\end{table*}

Similarly clear communities are also revealed in \textit{javax} network (\figref{resrwtech},~(b)). The core of the network consists of look-and-feel classes for different GUI components. Note that the majority of classes differ only in a small part of their name, which indicates the respective GUI component and look-and-feel implementation. In contrast to before, the community is not highly cohesive, as these classes are extensively used by, e.g., various GUI components. The latter in fact appear within the largest link-pattern community, which is thus strongly dependent on the former. Note also that the latter link-pattern community consists of almost all GUI components of Java, although they reside in various packages and their names (i.e., functions) differ substantially. For more details on community structures of both technological networks see~\tblref{resrwtech}.

Despite mostly qualitative analysis, general propagation algorithms indeed reveal significant community structures within these technological networks, while the communities can also be related to particularly clear functional roles. Obviously, the latter could not be detected under the classical framework of merely cohesive modules. Note also that the proposed algorithms do not only partition the underlying software systems, as in the case of classical community detection, but also reveal important dependencies among different subsystems that would otherwise remain concealed. It ought to be mentioned that we have previously conjectured the existence of other modules besides classical communities in software networks~\cite{SB11d}.

\begin{figure}[t]
\centering
\includegraphics[width=0.80\columnwidth]{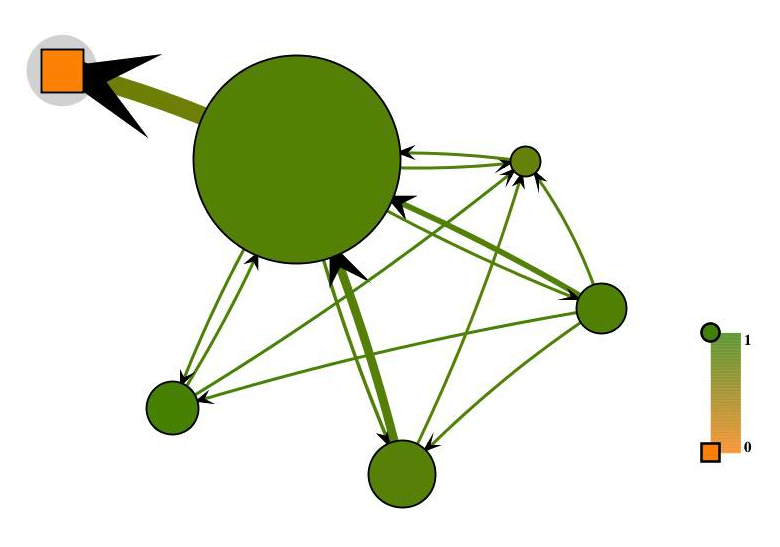}
\caption{\label{fig_resrwinf}(Color~online)~Community structure of \textit{amazon} information network revealed with \gpac. Edge directions were not considered by the algorithm.}
\end{figure}

Next, we analyze the community structure of \textit{amazon} information network that represents a small sample of Amazon web graph (\tblref{resrw}). The revealed network structure can be seen in~\figref{resrwinf}. Due to the size of the network and the nature of the domain, an exact analysis of extracted communities could not be conducted. Still, in the following, we discuss the main properties and highlight some interesting observations.

A large number of nodes is classified into dense core of the network ($1381$ nodes), however, the algorithm also reveals five well defined communities in the periphery (with $300$ nodes on average). Thus, as one could anticipate, the extracted partition rather accurately coincides with the core-periphery structure~\cite{LLDM09} that is commonly found in information networks~\cite{LLDM08,LLDM09}. For reference, value of $\delta_l$ for the core equals $0.86$, and is $0$ for the only link-pattern community. Communities in periphery exhibit $0.86$ on average.

\footnotetext[2]{This can be determined by the occurrence of '11091801' within the URL of the respective web page.}

We have analyzed the link-pattern community in great\-er detail and observed that the majority of its nodes correspond to web pages on musical instruments\footnotemark[2] sold on Amazon. In particular, $231$ of $288$ nodes represent web pages on various instruments, while each page corresponds to a different brand (e.g., Yamaha).  What makes the community particularly significant is the fact that only one of other $2591$ nodes in the network also represents a web page on musical instruments (the latter is the node connected to all nodes in the respective community). Hence, the algorithm manages to extract a meaningful link-pattern community from the core of the network, while the community is not only exhaustive but also rather clear.

Observe that link-density communities generally more strongly connect towards the core of the network, whereas, in the case of link-pattern community, the connection is significantly stronger in the direction from the core. As the network was treated as undirected, the latter cannot be considered as an artifact of the algorithm. The revealed pattern could in this context imply that nodes in link-pattern communities provide important content (i.e., authority nodes~\cite{Kle99}), while hub nodes~\cite{Kle99} reside mainly in link-destiny communities. Again, the occurrence of different types of communities can be related to a form of network heterogeneity (i.e., edge directions).

\begin{figure}[b]
\centering
\includegraphics[width=0.80\columnwidth]{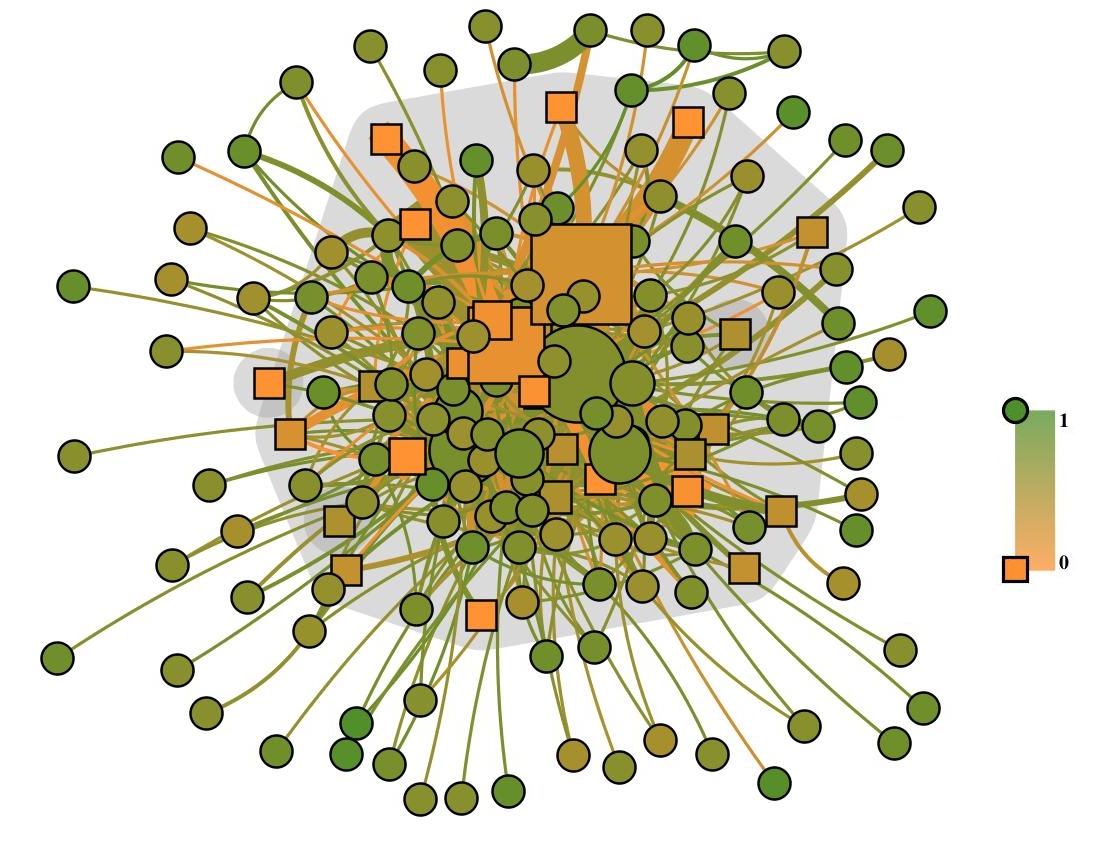}
\caption{\label{fig_resrwbio}(Color~online)~Community structure of \textit{protein} biological network revealed with \gpac.}
\end{figure}

For a complete analysis, we also apply the algorithm to an example of a biological network (that is also heterogeneous by definition). In particular, we analyze \textit{protein} network that represents protein-protein interactions of yeast Saccharomyces cerevisiae (\tblref{resrw}). The revealed community structure appears in~\figref{resrwbio}, while detailed description of communities is omitted. Observe that the algorithm reveals a large number of clear link-density and link-pattern communities of various sizes ($171$ communities of $2$ to $127$ nodes), while both exist in the interior and the periphery of the network. Different types of communities are combined in complex configurations (shaded regions in~\figref{resrwbio}), which, as in the examples above, suggests that link-pattern communities, similarly as link-density counterparts, are ubiquitous in real-world networks.

\begin{figure}[t]
\centering
\includegraphics[width=0.80\columnwidth]{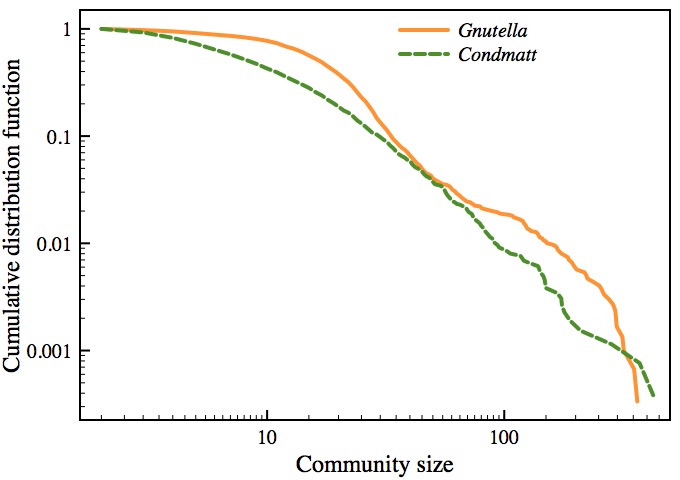}
\caption{\label{fig_resrwsize}(Color~online)~Cumulative size distributions of community structures revealed with \gpac~in \textit{gnutella} information and \textit{condmatt} social networks.}
\end{figure}

Last, we also analyze community structures revealed with \gpac~in two remaining networks in~\tblref{resrw}. More precisely, we consider \textit{gnutella} information network of peer-to-peer communications within Gnutella file sharing, and \textit{condmatt} social network representing scientific author collaborations extracted from Condensed Matter archive. While the former can be characterized by a unique dissassortative behavior, the latter is in fact a prominent example of assortative mixing, and thus classical community structure. Indeed, more than $92$ percent of the nodes in \textit{gnutella} network are classified into $2670$ link-pattern communities, while, on the contrary, almost $85$ percent of the nodes in \textit{condmatt} network reside in $2100$ classical link-density modules ($\delta_l$ equals $0.30$ and $0.64$ on average, respectively). \figref{resrwsize}~shows also cumulative community size distributions for both networks. Although distribution for \textit{condmatt} network appears to be power-law for the most part, as commonly observed in classical community detection~\cite{RCCLP04,PDFV05}, the latter does not hold for \textit{gnutella} network. In particular, communities most distinctively exists on two scales with tens and hundreds of nodes, which provides some evidence that link-pattern communities might reflect in disassortative mixing by degree (\secref{ldpc}).

%%%%%%%%%%%%%%%%%%%%%%%%%%%%%%%

\section{\label{sec_conc}Conclusions}
The paper proposes a balanced propagation based algorithm for detection of arbitrary network modules, ranging from classical cohesive (link-density) communities to more general link-pattern communities. The proposed algorithm was first validated on synthetic benchmark networks with community structure, and also on random networks. It was then further applied to different social, technological, information and biological networks, where it indeed reveals significant (composites of) link-density and link-pattern communities. In the case of larger real-world networks, the proposed algorithm more accurately detects the true communities than a state-of-the-art algorithm, while, in contrast to other approaches proposed in the literature, it does not require some prior knowledge of the true network structure. The latter is in fact crucial for the analysis of large real-world networks~\cite{KN11a}.

Heterogeneity appears to be a necessary condition for the network to reveal both link-density and link-pattern communities. However, although often not apparent at first sight, most real-world networks are in fact heterogeneous by nature. Qualitative results on real-world networks further imply that link-pattern communities, similarly as link-density counterparts, appear ubiquitous in nature and technology. Moreover, link-pattern communities are also commonly combined with classical modules into complex configurations, thus, different types of communities should not be analyzed independently. A generative model or measure for a general community structure of real-world networks would be of great benefit. It ought to be mentioned that the existence of link-pattern communities in real-world networks has implications in numerous other fields of network science (e.g., dynamic processes). 

The analysis in the paper does not directly imply which common properties of real-world networks one can expect under link-density or link-pattern regime. However, further work demonstrates that most significant link-pattern communities are revealed in regions with low values of clustering coefficients~\cite{WS98,SV05}, while just the opposite holds for classical modules. Furthermore, link-pattern communities may be the origin of degree disassortativity observed in various real-world networks~\cite{New02,NP03}, while they also commonly contradict the small-world phenomena~\cite{WS98}. Hence, different network properties seem to be governed by the same underlying principle~\cite{FFGP11}, which represents a prominent direction for future research.

%%%%%%%%%%%%%%%%%%%%%%%%%%%%%%%%%%%%%%%%%%%%%%%%%%%%%%%%%%%%

\begin{acknowledgement}
This work has been supported by the Slovene Research Agency ARRS within Research Program No. P2-0359.
\end{acknowledgement}

%%%%%%%%%%%%%%%%%%%%%%%%%%%%%%%

\bibliographystyle{epj}

%%%%%%%%%%%%%%%%%%%%%%%%%%%%%%%%%%%%%%%%%%%%%%%%%%%%%%%%%%%%

\end{document}